# Microbial Origin of Excess Greenhouse Gases in Glacial Ice

H. C. Tung,[1] N. E. Bramall,[2] and P. B. Price[2*]

University of California, Berkeley, CA 94720

## Abstract

We report the discovery of methanogenic archaea that account for abrupt 10× increases in methane concentration found by E. Brook at depths of 2954 and 3036 m in the GISP2 (Greenland Ice Sheet Project 2) ice core. The total microbial concentration we measured with direct cell counts tracks the excesses of methanogens that we identified by their F420 fluorescence. The highly localized (<1 m thick) layers of methanogens suggest flow-induced mixing of layers of microbe-laden anaerobic basal ice with glacial ice. The metabolic rate we found for microbes at 2954 and 3036 m lies roughly on the Arrhenius line for microbes imprisoned in rock, sediment, and basal ice. Equating the loss rate of methane recently discovered in the Martian atmosphere to the production rate by possible methanogens, we estimate that their Martian habitat would be at a temperature of ~0ºC and that the concentration, if uniformly distributed in a 100-m-thick layer, would be ~0.04 cell cm$^{-3}$.

[1] Dept. of Environmental Science, Policy, and Management
[2] Dept. of Physics, University of California, Berkeley, CA 94720, USA
* To whom correspondence should be addressed. E-mail: bprice@berkeley.edu

**One-sentence summary**: Abrupt 10× increases in methane concentration in thin (<1 m) layers of a Greenland ice core are found to be produced by methanogens and to correlate with 10× increases in total microbial cells in the same layers.

Studies of microbial life in cold environments on Earth help us to understand how life could have arisen on other planets. Bacteria and archaea have been found in all subfreezing terrestrial environments (1-5). Studies of terrestrial psychrophiles and psychrotrophs have involved extraction and examination of cells from water, ice, or permafrost. Instruments to look for chemical evidence of extant or extinct life in the Martian subsurface are being developed for future missions, and eventually Mars samples will be returned to Earth for study. Ideally, remote selection of such samples should be guided by molecular signatures of microbial life rather than just by surface morphology or rock type.

In a study of $N_2O$ in portions of the Vostok ice core, Sowers found a 30% excess at a depth corresponding to the penultimate glacial maximum (~135 kiloyears ago), where excess bacterial counts and dust had also been found, and he suggested that the $N_2O$ had been produced *in-situ* by nitrifying bacteria (6). Flückiger *et al.* have found excess $N_2O$ at several depths in the GRIP (Greenland Ice Core Project) ice core (7) and the NGRIP (north GRIP) ice core (8) and have dismissed them as unknown "artifacts" without looking for excess microbes in the ice at those depths. In our recent study (9) of GISP2 basal ice containing up to ~1 wt% silt grains, we found very high concentrations of microbes whose *in-situ* metabolism accounted for huge excesses of $CO_2$ and $CH_4$ found at the same depths (10). The silty, basal ice originated in glacial abrasion of a frozen swamp some $3 \times 10^5$ years ago (9).

Figure 1 shows data of Brook (11,12) on the concentration of methane as a function of depth in samples from the GISP2 ice core taken at sparse intervals of ~4 to ~10 m in depth. The present work began when we noticed that his values at 2954 and 3036 m stood out as being an order of magnitude higher than at any other depths he studied (116 to 3038 m). We report here our discovery of methanogenic archaea and other microbial genera at those two depths in clear ice, and that their metabolic rates corresponded to a cellular carbon turnover time of ~300,000 yrs at an ice temperature -11ºC.

From the National Ice Core Laboratory we obtained samples of GISP2 ice at depths ~2954 and ~3036 m, adjacent, within a few cm, to the locations of the samples in which Brook found excess methane. In addition, we obtained ice samples at eight other depths from 150 to 3000 m as a control. We sterilized our equipment by high-temperature baking. In a sterile laminar flow hood, we removed >30% of the exterior of each GISP2 ice sample by rinsing it sequentially with 10M HCl, DNAse- and RNAse-free sterile water, 10 M NaOH, and a final rinse in sterile water in order to eliminate any contamination that might have been introduced during the drilling process or subsequent handling of the samples. This very stringent treatment guaranteed that not only cells but also bacterial spores and nucleic acid were removed from the exterior and destroyed. To test our methods, we made numerous ice samples with sterile water, coated them with various types of bacteria and bacterial spores, subjected them to the same sterilizing procedure we used with the GISP2 sample, and verified, with direct counting methods and culturing, that none of the microbes coated on the ice surfaces made it into the final test sample. This test was verified many times. We processed the GISP2 ice samples together with control samples in a double-blind manner. If, at any stage of our study, one of the control samples showed signs of contamination, we disregarded measurements on accompanying samples.

Microbes in the fluid used for drilling ice cores have been shown to contaminate the outer portions of cores on which biological studies have been done. Christner *et al.* (13) evaluated microbial contamination of the Lake Vostok core by measuring the radial gradient of microbial concentration in a core section. Following their procedure, we measured the microbial concentration as a function of radial distance from the outer surface in a GISP2 sample from a depth of 500 m. Our direct cell counts, plotted in Fig. 2, showed that some of the outermost 3.8 cm portion had been contaminated. Consequently, we took all of our ice samples from the innermost 2.5 cm extending to the center of the core where the microbial concentration was independent of radial distance.

To count total concentrations of cells, we melted an interior sample at each depth, immediately filtered it through a 0.015 μm Nuclepore filter, stained cells with up to 20 μM of Syto 23, and used a Zeiss Axiovert epifluorescence microscope with a 100X objective. Using the same microscope, we counted methanogens in a melted, unstained, interior sample by viewing the blue-green autofluorescence of the F420 co-enzyme that is accepted as a unique signature of methanogens (14). Using excitation at 420 nm and a barrier filter with cutoff at 460 nm, we observed this fluorescence in a small fraction of cells in the band ~460-490 nm. To eliminate possible interference due to autofluorescence of micron-size mineral grains that might accompany the microbes, we first exposed cells of the methanogen *Methanococcus jannaschii* to 420 nm light at high intensity until their F420 fluorescence was destroyed by photobleaching. Then, we subjected mineral grains filtered from melted GISP2 ice to the same photobleaching process. The results showed that mineral grains from GISP2 were far more resistant to photobleaching than were methanogens. Moreover, the minerals tended to fluoresce over a broad spectrum of excitation wavelengths, whereas F420 fluoresced only at 420nm excitation. We used these facts to distinguish F420 signals from possible fluorescence of mineral grains. Although F420 has been detected in cells of several members of all three domains (15), where it serves a variety of roles, it is abundant only in methanogens. In nonmethanogens its concentration is typically lower by several orders of magnitude (16), and its detection requires purification. Only in methanogens is its autofluorescence visible.

Figure 3 shows the concentrations of Syto23-stained cells (top) and of cells detected, in a double-blind manner, by their F420 autofluorescence (bottom) as a function of depth. At depths ≤2000 m the concentrations of methanogens and of Syto23-stained cells were <90/ml (based on null counts in 50 ml) and ~$2 \times 10^4$/ml respectively. We found methanogens and large excesses of Syto23-stained cells at 2954, 3000, and 3036 m. In addition, at 2238 m we found a high concentration of Syto23-stained cells but no methanogens (< 90/ml). Since Brook did not measure methane at 2238 m, the null result for methanogens is not significant. Brook measured methane at 3000 m but did not find an excess.

The correspondence of Brook's high methane concentrations with our high concentrations of microbes, and especially of methanogens, at depths 2954 and 3036 m strongly supports our hypothesis that methanogenic metabolism accounts for the anomalously high methane in GISP2 ice. The large excess of nonmethanogenic microbes at those two depths provides independent evidence that the excess methane was microbial. The absence of excess methane where we saw methanogens at ~3000 m suggests that the excess methane and methanogens were localized in an interval less than ~1 m deep that did not exactly correspond with the depth of Brook's sample. We conclude that at least three thin layers of ice with high-methane and high-microbial content were excavated from basal, microbe-rich ice by turbulent glacial flow during buildup of the ice sheet. We think it very unlikely that methanogens, which are strict anaerobes, could have survived several days in the atmosphere in transit to the Greenland ice.

In their measurements of $N_2O$ at 1-m depth intervals in the NGRIP core, Flückiger et al. (8) found ten anomalously high values localized in isolated 1-m intervals between 2100 and 2240 m and four high values between 2540 and 2590 m. This rather common occurrence of excess $N_2O$ of possible biogenic origin suggests to us that, if methane were to be measured at 1-m depth intervals in the GISP2 core, additional layers of excavated basal ice rich in microbes would be discovered.

Price and Sowers (17) recently showed that the metabolic rates of communities of microbes imprisoned in ice and other solid media can be calculated from the concentrations of trapped gas that they produced during a time $t$ at an absolute temperature $T$. Equation (1) gives the rate $R(T)$, defined as the fractional rate of turnover of carbon per cell per year:

$$R(T) = Y_j(T)/n_j\, m_j\, t \qquad (1)$$

where $Y_j(T)$ is the concentration of biogenic gas of type $j$ at ice temperature $T$ (18); $n_j$ and $m_j$ are concentration and mean mass per microbe of type $j$; and $t$ is the retention time of the gas in the ice. We used this expression to calculate metabolic rates for production of $CH_4$ and $CO_2$ (19). We took $m_j$ = 19 fg, based on our measurements of the typical size of cells in the silty ice at the bottom of the GISP2 core (9). Figure 4 shows our results, together with results from Price and Sowers (17), on an Arrhenius plot of log $R$ vs $1/T$. The results of Price and Sowers are shown as solid squares. The points labeled by solid diamonds are our results for methanogenic metabolism in clear ice at 2954 and 3036 m (Fig. 3), for which gas retention ages are taken to be 143 and 163 kyr (20). Arrows indicate that the rates might be overestimated, if the scanning efficiency for detecting methanogens were less than 100% due to a weaker than normal F420 fluorescence of some species. Such would be the case if some methanogens had a lower than typical concentration of F420 molecules or if the F420 were reduced (14) or had been in a low-pH environment (14) such as acidic veins (21). In calculating metabolic rates for $CO_2$ production (points shown as open triangles), we assumed that cells imaged with Syto-23 fluorescence were mainly anaerobic $CO_2$ producers such as Fe-reducers and sulfate reducers, and that the ratio of $CO_2$ to $CH_4$ concentrations was 20, as was found in the basal ice (10).

The dashed line in Fig. 4 is an extrapolation of an Arrhenius line for spontaneous racemization of aspartic acid in microbes from Siberian permafrost, measured at temperatures 100ºC to 145ºC (22). Rates for racemization of the other amino acids and for depurination of DNA (23) are an order of magnitude lower and can be neglected. The error in racemization rate extrapolated into the interval 30ºC to -40ºC is about a factor two. Assuming that the activation energy is constant from 145ºC to -40ºC, the approximate agreement of the racemization rate with the metabolic rate for survival of immobilized microbes in ice and shale suggests that the rate of repair of macromolecular damage in living microbes may be the same as the rate of spontaneous damage (17). If so, a fascinating picture emerges: microbes adapted to conditions of temperature, darkness, pH, pressure, redox value, nutrient source, and desiccation in the environment in which they are trapped may survive for millions of years, utilizing the limited source of energy and nutrients not to grow but just to repair macromolecular damage at the same rate at which it spontaneously occurs. This strategy of "repair as you go" differs from that of endospores, which do not repair damage until after conditions improve enough for them to germinate.

Visual inspection of the GRIP and GISP2 ice cores by others (24) provided evidence that, at depths below ~2450 m, ice flow caused frequent overturning or flow-induced mixing (10). Our results support that view: the high concentrations of microbes localized at 2954, 3000, and 3036 m likely resulted from interleaving of basal ice with clear ice on a depth scale finer than ~1 m, at least in the lowest hundred meters or so. Our discovery of methanogens that metabolized in the ice indicates that, as strict anaerobes, they must have come from below since they could not have survived atmospheric transport.

Our measurements of the temperature-dependent metabolic rate of confined microbes have an interesting application to life on other planets. Krasnopolsky et al. (25) and Formisano et al. (26) recently detected ~10 ppb of methane in the Martian atmosphere. Krasnopolsky et al. estimated a lifetime of ~340 yrs for destruction of methane in the atmosphere by solar UV photolysis and a corresponding loss rate of 270 tons of methane per year. They calculated that abiotic production mechanisms cannot readily offset such a loss rate, and they concluded that ongoing methanogenesis by subsurface methanogens is a plausible source. Knowing the temperature-dependent metabolic rate for methanogenesis (our Fig. 4), we can calculate the required number of Martian methanogens. Setting the steady-state metabolic rate of production of methane equal to the estimated loss rate of $1/340 = 3 \times 10^{-3}$ yr$^{-1}$, we find, using the rate given by the solid diamonds in Fig. 4, that the temperature of a Martian methanogenic habitat would be ~0°C. From (27), we find that the Martian subsurface temperature reaches 0°C at a depth between 150 m and 8 km, depending on soil thermal conductivity, and that the time for the methane to diffuse to the surface would be 15 yr if from 150 m or 30,000 yr if from 8 km.

For a uniform distribution of methanogens over the Martian subsurface in a thickness interval $\delta z$ where the temperature is 0°C, and for a mass $m$ of carbon per cell, we estimate that their concentration in the spherical subsurface shell would be

$$N \text{ [cm}^{-3}\text{]} = 0.04 \ (10^2 \text{ m}/\delta z) \ (15 \text{ fg}/m) \qquad (2)$$

Although it may be possible with a sensitive fluorimeter to detect F420 autofluorescence of 0.04 cell cm$^{-3}$, it is unlikely that an instrument can be deployed to a depth of 150 m to 8 km in the foreseeable future. A more practicable strategy would be to use a surface rover to search for sites of above-average methane concentrations (26).

**References and Notes**

1. See references in J. C. Priscu and B. Christner, in *Microbial Diversity and Bioprospecting*, A. T. Bull, Ed. (ASM Press, Washington, D. C., 2004), pp. 130-145.

2. E. Gaidos *et al., Astrobiology* **4**, 327 (2004).

3. S. A. Bulat *et al., Intern. J. Astrobiol.* **3**, 1 (2004).

4. P. P. Sheridan *et al., Appl. Environ. Microbiol.* **69**, 2153 (2003).

5. V. I. Miteva *et al., Appl. Environ. Microbiol.* **70**, 202 (2004).

6. T. Sowers, *J. Geophys. Res.* **106**, 31903 (2001).

7. J. Flückiger *et al., Science* **285**, 227 (1999).

8. J. Flückiger *et al.*, *Global Biogeochem. Cycles* **18**, GB1020 (2004).

9. C. Tung, P. B. Price, and N. Bramall, submitted to *Astrobiology* (2005).

10. R. Souchez, M. Lemmens, and J. Chappellaz, *Geophys. Res. Lett.* 22, 41 (1995).

11. E. Brook, NOAA Geophysical Data Center, www.ngdc.noaa.gov/paleo/paleo.html.

12. For depths 2808 to 3038 m, Ed Brook provided us with an unpublished table of his methane measurements.

13. B. C. Christner *et al.*, *Icarus* **172**, 572 (2005).

14. A. A. DiMarco, T. A. Bobik, and R. S. Wolfe, *Annu. Rev. Biochem.* **59**, 355 (1990).

15. E. Purwantini and L. Daniels, *J. Bac.* **180**, 2212 (1998).

16. X. L. Lin and R. H. White, *J. Bac.* **168**, 444 (1986).

17. P. B. Price and T. Sowers, *Proc. Natl. Acad. Sci. USA* **101**, 4631 (2004).

18. R. B. Alley *et al.*, GISP2 Ice Core Temperature and Accumulation Data. IGBP PAGES, World Data Center for Paleoclimatology Data Contribution Series #2004-013. NOAA/NGDC Paleoclimatology Program, Boulder CO, USA (2004).

19. We assumed that methanogens and $CO_2$-metabolizers are located in a network of liquid veins along junctions where three ice grains intersect (21), that the reaction for methanogens is $4HCOOH \rightarrow CH_4 + 3CO_2 + 2H_2O$, that a typical reaction for $CO_2$-metabolizers is $2Fe^+ + HCOOH \rightarrow 2Fe^{2+} + 2H^+ + CO_2$, and that the source of the formate is the silty ice at the base of the glacier (9).

20. D. A. Meese *et al.*, *J. Geophys. Res.* **102** (C12), 26411 (1997).

21. P. B. Price, *Proc. Natl. Acad. Sci. USA* **97**, 1247 (2000).

22. K. L. Brinton et al., *Astrobiology* **2**, 77 (2002).

23. T. Lindahl and N. Nyberg, *Biochemistry* **11**, 3610 (1972).

24. R. B. Alley *et al.*, *Nature* **373**, 393 (1995).

25. V. A. Krasnopolsky *et al.*, *Icarus* **172**, 537 (2004).

26. V. Formisano et al., *Science* **306**, 1758 (2004).

27. M. T. Mellon and R. J. Phillips, *J. Geophys. Res.* **106** (E10) 23165 (2001).

28. We thank Ed Brook for providing us with his unpublished methane data, Eric Craven for cutting and shipping the GISP2 samples, Boonchai Boonyaratanakornkit for providing the *M. jannaschii* cells, and NSF Office of Polar Programs for grants OPP-0085400 and REU-0343999.

**Figure Captions**

1. Methane concentration as a function of depth in the GISP2 ice core, measured by Brook (11,12). For clarity, only his data at depths greater than 2700 m are shown. Arrows indicate anomalously high values.

2. Microbial concentration as a function of radial distance from the surface to the center of the GISP2 ice core at a depth 500 m, showing evidence of contamination of the outer region.

3. Concentrations of cells stained with Syto 23 (top) and of methanogenic cells imaged in F420 autofluorescence (bottom). Error bars are counting statistics.

4. Arrhenius plot of metabolic rate as a function of $1/T$, with $T$ in Kelvins, for microbial communities imprisoned in ice, rock, and sea sediment. Metabolic products are shown in parentheses. Solid squares are from (17); solid diamonds are for methanogens; open triangles are for $CO_2$ producers. See text for implications of the near equality of spontaneous macromolecular damage and of metabolism.

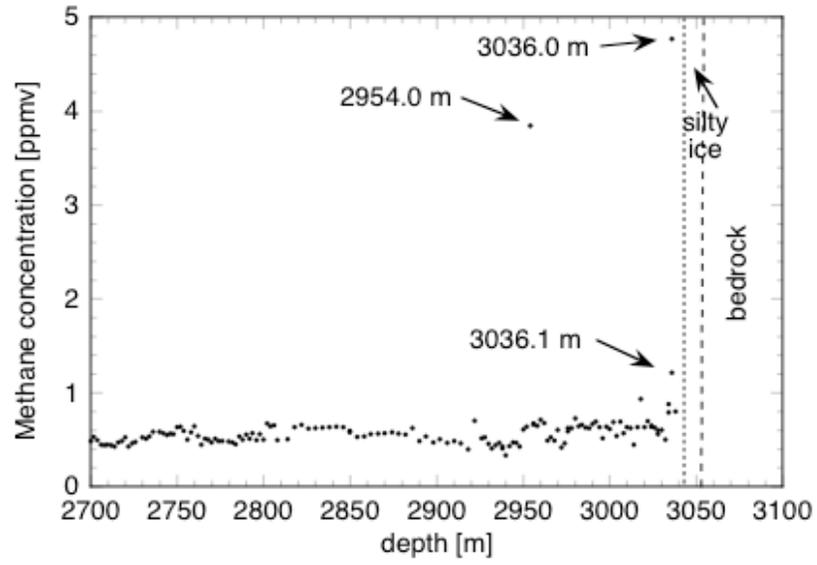

Figure 1

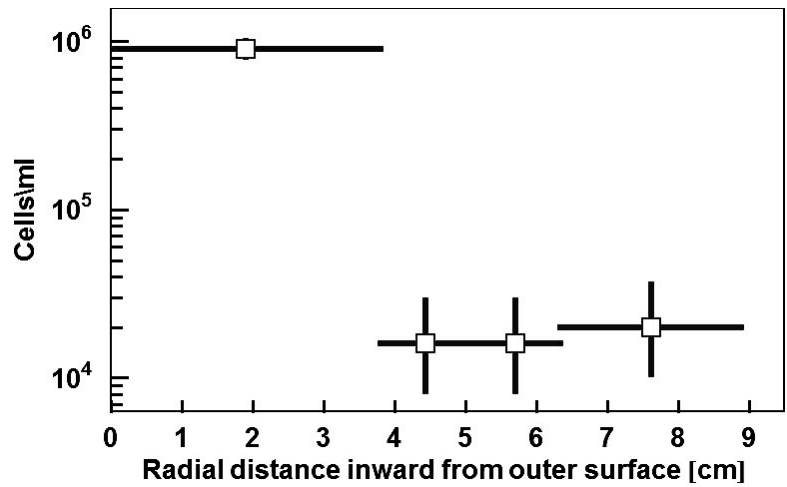

Figure 2

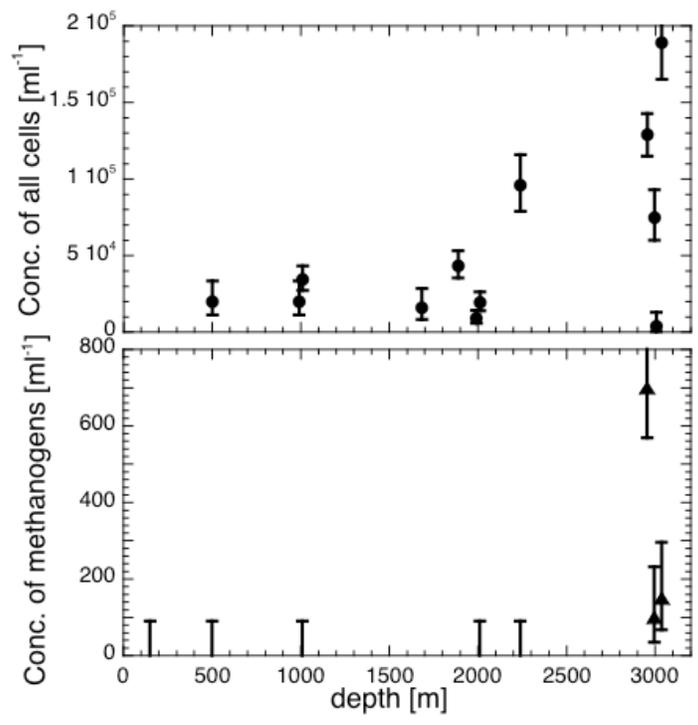

Figure 3

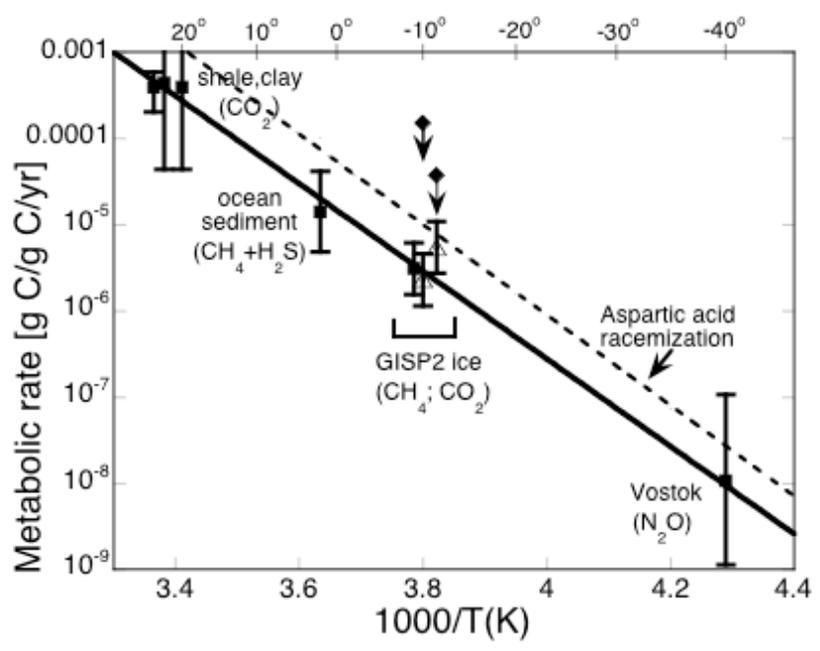

Figure 4